\documentstyle[12pt]{article}
\newcommand{\beq}{\begin{equation}}
\newcommand{\eeq}{\end{equation}}
\newcommand{\bea}{\begin{eqnarray}}
\newcommand{\eea}{\end{eqnarray}}

\newcommand{\goto}{\rightarrow}

\begin{document}
\setcounter{page}{0}
\topmargin 0pt
\oddsidemargin 5mm
\renewcommand{\thefootnote}{\fnsymbol{footnote}}
\newpage
\setcounter{page}{0}
\begin{titlepage}

\begin{flushright}
QMW-PH-97-5\\
OU-TP-97-11P\\
{\bf hep-th/9703028}\\
 {\it February 1997}
\end{flushright}
\vspace{0.5cm}
\begin{center}
{\Large {\bf Knizhnik-Zamolodchikov-type equations for gauged WZNW models}} %%@
\\
\vspace{1.8cm}
\vspace{0.5cm}
{Ian I. Kogan$~^{1,}$\footnote{e-mail:
i.kogan1@physics.oxford.ac.uk}, Alex
Lewis$~^{1,}$\footnote{e-mail:
a.lewis1@physics.oxford.ac.uk} and  Oleg A.
Soloviev$~^{2,}$
\footnote{e-mail: O.A.Soloviev@QMW.AC.UK}} \\
\vspace{0.5cm}
{\em$~^1$Department of Physics, University of Oxford\\
1 Keble Road, Oxford, OX1 3NP, United Kingdom}\\
{\em$~^2$ Physics Department, Queen Mary and Westfield College, \\
Mile End Road, London E1 4NS, United Kingdom}\\
\vspace{0.5cm}
\renewcommand{\thefootnote}{\arabic{footnote}}
\setcounter{footnote}{0}
\begin{abstract}
{We study correlation functions of coset constructions by utilizing the %%@
method of gauge dressing. As an example we apply this method to the minimal %%@
models and to the Witten 2D black hole. We exhibit a striking similarity %%@
between the latter and the gravitational dressing. In particular, we look %%@
for logarithmic operators in the 2D black hole.
% and discuss their importance in physics. 
%By analogy with the gravitational dressing, we analyze the effect of the %%@
%%@
%gauge dressing on renormalization group flows.
}
\end{abstract}
\vspace{0.5cm}
%\centerline{July 1996}
 \end{center}
\end{titlepage}
\newpage
%******************************************************************
\section{Introduction}

Gauged Wess-Zumino-Novikov-Witten (WZNW) models \cite{gawedzki}-\cite{BH} %%@
belong to the type of 2D quantum field theories whose correlation functions %%@
can be studied systematically in a way which is similar to the Knizhnik-%%@
Zamolodchikov (KZ) approach to ordinary WZNW models \cite{Knizhnik}. This %%@
approach being based upon clear physical principles has proved powerful in %%@
computation of correlators of WZNW models. Though one has to point out that %%@
there are a number of different rather algebraic algorithms one can make %%@
use of to find WZNW correlation functions, e.g. %%@
\cite{Dotsenko},\cite{Naculich}. 

A central element of the KZ method is a differential equation which is now %%@
called KZ equation \cite{Knizhnik}. Correlation functions of the WZNW model %%@
turn out to be solutions of this equation. Since gauged WZNW models are %%@
closely related to ordinary WZNW models, it is quite natural to conjecture %%@
that correlators of the gauged WZNW model have to obey a certain gauge %%@
generalization of the KZ equation. In particular, Polyakov \cite{Polyakov1} %%@
has shown that correlation functions of gauged fermions do satisfy a KZ-%%@
type equation. At the same time, Witten has proved that free fermions are %%@
equivalent to the WZNW model at level one \cite{Witten}. It is clear that %%@
the given equivalence has to hold for the gauged models as well. Hence, %%@
correlation functions of the gauged WZNW model at level one has to satisfy %%@
the same KZ-type equation of the gauged fermions. In \cite{Alex}, we have %%@
derived a differential equation for correlators of the general gauged WZNW %%@
model following exactly this intuition. A different generalization of the %%@
KZ equation has been obtained in \cite{Halpern1}. It is based on the so-%%@
called affine-Virasoro construction whose conformal algebra coincides with %%@
the one of the stress tensor of the gauged WZNW model (for review see %%@
\cite{Halpern}). 

Gauged WZNW models have been under extensive consideration for quite a long %%@
time mainly because they describe coset constructions \cite{Goddard} which %%@
play an important role in string
theory and
statistical physics. A recent discussion of these 
models is \cite{witten}.
While algebraic properties of cosets have been quite well understood, much %%@
less is known about their correlation functions (except, perhaps, the case %%@
of the minimal models). The coset correlation functions are crucial for %%@
understanding of dynamical phenomena in theories such for example as two %%@
dimensional black holes \cite{Witten1}.

Polyakov's approach to the gauge coupling also has been used to study %%@
gravitationally dressed correlation functions of general quantum field %%@
theories \cite{polyakov},\cite{kpz},\cite{Klebanov},\cite{bilal} (see also %%@
\cite{chams}). Both gravitational and gauge interactions have many features %%@
in common. In fact, as we shall show in the present paper, the %%@
gravitational dressing can be equivalently described as the gauge dressing. %%@
What seems to have been missed so far form the discussion is the mix of %%@
gravitational and gauge dressings. This issue will be addressed in the %%@
present paper.

In the present paper we would like to present all the details of our %%@
derivation of the gauged KZ equation \cite{Alex} which is the subject of %%@
section 2. In section 3, we discuss how the known correlation functions of %%@
the minimal models emerge as solutions of our equation. In section 4, we %%@
carry on to look for the solutions of our equation in two other interesting %%@
cases of parafermions and  Witten's 2D black hole. We argue that the latter %%@
gives rise to the differential equation which is similar to the equation %%@
describing the gravitational dressing. In section 5, we derive an equation %%@
which describes the mix of gravitational and gauge dressings. Section 6 %%@
contains conclusion and discussion of our results.

\section{Gauge dressing of the Knizhnik-Zamolodchikov equation}

A large class of 2D conformal
field theories is described by gauged WZNW models with the following
action ( we use here the same  normalization as in \cite{witten})
\begin{equation}
S(g,A)=S_{WZNW}(g)~+~{k\over2\pi}\int d^2z\mbox{Tr}\left[
Ag^{-1}\bar\partial g - \bar A\partial g g^{-1}
+Ag^{-1}\bar Ag
{}~-~
A\bar
A\right],\label{action}\end{equation}
where
\begin{equation}
S_{WZNW}(g)={k\over8\pi}\int d^2z~\mbox{Tr}g^{-1}\partial^\mu
gg^{-1}\partial_\mu g~+~{ik\over12\pi}\int
d^3z~\mbox{Tr}g^{-1}dg\wedge
g^{-1}dg\wedge g^{-1}dg\label{wznw}\end{equation}
and $g\in G$, $A,~\bar A$ are the gauge fields taking values in the
algebra
${\cal H}$ of the diagonal group of the direct product $H\times H$,
$H\in G$.

It has become usual to study gauged WZNW models with the BRST method
\cite{Karabali}. However, this method is not very much of help in
computing
correlation functions, though, in principle, the free field
realization of
these theories allows one to calculate correlators of BRST invariant
operators.
We shall pursue a different approach which is parallel to the
analysis of the
gravitational dressing of 2D field theories.

Our starting point are the equations of motion of the gauged WZNW
model:
\begin{eqnarray}
\bar\nabla(\nabla gg^{-1})&=&0,\nonumber\\ &\label{eqmotion}&\\
\bar\partial A~-~\partial\bar A~+~[A,\bar
A]&=&0,\nonumber\end{eqnarray}
where
\begin{equation}
\bar\nabla=\bar\partial~+~\bar
A,~~~~~~\nabla=\partial~+~A.\label{nabla}
\end{equation}

Under the gauge symmetry, the WZNW primary fields $\Phi_i$ and the
gauge
fields  $A,~\bar A$ transform respectively as follows
\begin{eqnarray}
\delta\Phi_i&=&\epsilon^a(t^a_i+\bar t^a_i)\Phi_i,\nonumber\\
\delta A&=&-\partial\epsilon-[\epsilon,A],\label{gtrans}\\
\delta\bar A&=&-\bar\partial\epsilon-[\epsilon,\bar
A],\nonumber\end{eqnarray}
where $t^a_i\in{\cal H}$.

In order to fix the gauge invariance, we impose the following
condition
\begin{equation}
\bar A=0.\label{newgauge}\end{equation}
The given gauge fixing gives rise to the corresponding Faddeev-Popov
ghosts
with the action
\begin{equation}
S_{ghost}=\int d^2z~\mbox{Tr}(b\partial
c).\label{ghosts}\end{equation}

In the gauge (\ref{newgauge}), the equations of motion take the
following form
\begin{eqnarray}
\bar\partial J&=&0,\nonumber\\ &\label{geqmotion}&\\
\bar\partial A&=&0,\nonumber\end{eqnarray}
where
\begin{equation}
J=-{k\over2}\partial
gg^{-1}~-~{k\over2}gAg^{-1}.\label{J}\end{equation}
Thus, $J$ is a holomorphic current in the gauge (\ref{newgauge}).
Moreover, it
has canonical commutation relations with the field $g$ and itself:
\begin{eqnarray}
\left\{J^a(w),g(z)\right\}&=&t^ag(z)\delta(w,z),\nonumber\\
&\label{canonical}&
\\
\left\{J^a(w),J^b(z)\right\}&=&f^{abc}
J^c(z)\delta(w,z)~+~k/2\delta^{a
b}\delta'(w,z).\nonumber\end{eqnarray}
The given commutators follow from the symplectic structure of the
gauged WZNW
model in the gauge (\ref{newgauge}). In this gauge, the field $A$
plays a role
of the parameter $v_0$ of the orbit of the affine group $\hat G$
\cite{Alekseev}. Therefore, the symplectic structure of the gauged
WZNW model
in the gauge (\ref{newgauge}) coincides with the symplectic structure
of the
original WZNW model \cite{Faddeev}.

There are residual symmetries which survive the gauge fixing
(\ref{newgauge}).
Under these symmetries the
fields $\Phi_i$ and the remaining gauge field $A$ transform according
to
\begin{eqnarray}
\tilde\delta\Phi_i&=&(\epsilon^A_L
t^A_i~+~\epsilon_R^a\bar t^a)\Phi_i,\nonumber\\&
\label{nonab}
\\
\tilde\delta A&=&-\partial\epsilon_R-[\epsilon_R,
A],\nonumber\end{eqnarray}
where the parameters $\epsilon_L$ and $\epsilon_R$ are
arbitrary
holomorphic functions,
\begin{equation}
\bar\partial\epsilon_{L,R}=0.
\label{tildeepsilon}\end{equation}
In eqs. (\ref{nonab}) the generators $t^A$ act on the left index of
$\Phi_i$,
whereas $\bar t^a$ act on the right index of $\Phi_i$. One can notice
that the
left residual group is extended to the whole group $G$, whereas the
right
residual group is still the subgroup $H$.

Eq. (\ref{J}) can be presented in the following form
\begin{equation}
\frac{1}{2}\partial g~+~\frac{\eta}{2}gA~+~\frac{1}{\kappa} Jg=0.
\label{maineq}\end{equation}
Here $\eta$ and $\kappa$ are some renormalization constants due to
regularization of the singular products $gA$ and $Jg$.

In order to compute $\eta$ and $\kappa$, we need to do a few things.
First of
all, we have to find how the gauge field $A$ acts on the fields
$\Phi_i$. To
this end, let us define dressed correlation functions
\begin{equation}
\langle\langle\cdot\cdot\cdot\rangle\rangle\equiv\int{\cal D}\bar
A{\cal
D}A\langle\cdot\cdot\cdot\rangle~\exp\left[-{k\over2\pi}\int
d^2z\mbox{Tr}\left\{\bar Ag^{-1}\partial g+A\bar\partial
gg^{-1}+Ag\bar Ag^{-1}~+~A\bar A\right\}\right]
,\label{dressing}\end{equation}
where $\langle\cdot\cdot\cdot\rangle$ is the correlation function
before
gauging. The latter is found as a solution to the KZ equation
\begin{equation}
\left\{
{1\over2}{\partial\over\partial z_i}~+~\sum^N_{j\ne
i}{t^A_it^A_j\over
k+c_V(G)}{1\over z_i-
z_j}\right\}\langle\Phi_1(z_1,\bar z_1)\Phi_2(z_2,
z_2)\cdot\cdot\cdot\Phi_N(z_N,\bar z_N)
\rangle=0.\label{knizhnik}\end{equation}
Here $\Phi_i$ are the primary fields of the WZNW model (\ref{wznw}),
$t^A_i$
are the representations of the generators of $G$ for the fields
$\Phi_i$,
\begin{equation}
c_V={f^{abc}f^{abc}\over\dim
G}.\label{cazimirs}\end{equation}

In the gauge (\ref{newgauge}), the dressed correlation functions
(\ref{dressing}) can be presented as follows
\begin{eqnarray}
\langle\langle\Phi_1(z_1,\bar z_1)\Phi_2(z_2,\bar
z_2)\cdot\cdot\cdot\Phi_N(z_N,\bar z_N)\rangle\rangle
=\int{\cal D}b{\cal D}c\exp(-S_{ghost})~\int{\cal D}
A\exp[-S_{eff}(
A)]\nonumber\\&\label{definition} &\\
\times\int{\cal D}g~\Phi_1(z_1,\bar z_1)\Phi_2(z_2,\bar
z_2)\cdot\cdot\cdot\Phi_N(z_N,\bar
z_N)~\exp[-\Gamma(g,A)],\nonumber\end{eqnarray}
where $S_{eff}(A)$ is the effective action of the field $A$ and
$\Gamma(g,A)$
is formally identical to the original gauged WZNW action in the gauge
(\ref{newgauge}). The
action $S_{eff}$
is non-local and can be obtained by integration of the following
variation
(which follows from the Wess-Zumino anomaly condition)
\begin{equation}
\partial{\delta S_{eff}\over\delta A^a}
{}~+~f^{abc}A^c{\delta S_{eff}\over\delta A^b}
=\tau\bar\partial A^a.\label{eff}\end{equation}
Here the constant $\tau$ is to be defined from the consistency
condition
of the gauge (\ref{newgauge}), which is
\begin{equation}
J_{tot}\equiv\delta Z/\delta\bar A^a=0,~~~~~~~a=1,2,...,\dim
H,\label{cond}\end{equation}
at $\bar A=0$. Here $Z$ is the partition function of the gauged WZNW
model.
Condition (\ref{cond}) amounts to the vanishing of the central charge
of the
affine current $J_{tot}$. This in turn means that $J_{tot}$ is a
first class
constraint \cite{Karabali}. In order to use this constraint, we need
to know
the OPE of $A$ with itself. This can be derived as follows. Let us
consider the
identity
\begin{equation}
\tau\langle\langle\bar\partial A( z) A(
z_1)\cdot\cdot\cdot
A( z_N)\rangle\rangle=\int{\cal D} A~ A(
z_1)\cdot\cdot\cdot A( z_N)\left[\partial{\delta
S_{eff}\over\delta
A^a( z)}~+~f^{abc}A^c(z){\delta
S_{eff}\over\delta
A^b( z)}\right]\mbox{e}^{-S_{eff}}.\label{equality}\end{equation}
Here we used relation (\ref{eff}). Integrating by parts in the path
integral, we arrive at the following formula
\begin{eqnarray}
\tau\langle\langle A^a( z) A^{a_1}( z_1)\cdot\cdot\cdot
A^{a_N}(
z_N)\rangle\rangle
={1\over2\pi i}\sum^N_{k=1}\{{-\delta^{aa_k}\over( z-
z_k)^2}\langle\langle\ A^{a_1}( z_1)\cdot\cdot\cdot\hat
A^{a_k}_k\cdot\cdot\cdot A^{a_N}(
z_N)\rangle\rangle
\nonumber\\&\label{formula}&\\
+{f^{aa_kb}\over z-z_k}\langle\langle\ A^{a_1}( z_1)\cdot\cdot\cdot
A^b_k\cdot\cdot\cdot A^{a_N}(
z_N)\rangle\rangle\},\nonumber\end{eqnarray}
where $\hat A_k$ means that the field $A(z_k)$ is removed from the
correlator.
In the derivation of the last equation we used the following identity
\begin{equation}
\bar\partial_{\bar z}{1\over z- z_k}=2\pi
i\delta^{(2)}(z-z_k).\label{delta}\end{equation}

{}From eq. (\ref{formula}) it follows that
\begin{equation}
\tau A^a(z)A^b(0)={1\over2\pi i}\left[-{\delta^{ab}\over z^2}
{}~+~{f^{abc}\over z}A^c(0)\right]~+~\mbox{reg}.
\label{gaugeope}\end{equation}
Along with condition (\ref{cond}), the equation (\ref{gaugeope})
gives the
expression for $\tau$
\begin{equation}
\tau={i(k+2c_V(H))\over4\pi}.\label{tau}\end{equation}

We proceed to derive the Ward identity associated with the residual
symmetry
(\ref{nonab}). The Ward identity comes  from the change of variables in
eq.
(\ref{definition}) under transformations (\ref{nonab}). We obtain the %%@
following relation
\begin{eqnarray}
\sum_{k=1}^N\bar t^a_k\delta(z,z_k)\langle\langle\Phi_1(z_1,\bar
z_1)\Phi_2(z_2,\bar
z_2)\cdot\cdot\cdot\Phi_N(z_N,\bar
z_N)\rangle\rangle\nonumber\\&\label{identity} &\\
+\tau\langle\langle\bar\partial_{\bar z}A^a(z)\Phi_1(z_1,\bar
z_1)\Phi_2(z_2,\bar
z_2)\cdot\cdot\cdot\Phi_N(z_N,\bar
z_N)\rangle\rangle=0.\nonumber\end{eqnarray}
This yields
\begin{eqnarray}
2\pi\tau\langle\langle A^a(z)\Phi_1(z_1,\bar
z_1)\Phi_2(z_2,\bar
z_2)\cdot\cdot\cdot\Phi_N(z_N,\bar
z_N)\rangle\rangle\nonumber\\&\label{Ward}
&\\
=i\sum_{k=1}^N{\bar t^a_k\over z- z_k}\langle\langle\Phi_1(z_1,\bar
z_1)\Phi_2(z_2,\bar z_2)\cdot\cdot\cdot\Phi_N(z_N,\bar
z_N)\rangle\rangle,\nonumber\end{eqnarray}
which in turn gives rise to the OPE between the gauge field $A^a$ and
$\Phi_i$
\begin{equation}
 {1\over2}A^a(z)\Phi_i(0)={1\over k+2c_V(H)}{\bar t^a_i\over
z}\Phi_i(0).\label{nonabope}\end{equation}

Now we are in a position to define the product $[A^a,\Phi_i]$.
Indeed,
we can
define this according to the following rule
\begin{equation}
 A^a(z)\Phi_i(z,\bar z)=\oint{d\zeta\over2\pi i}{
A^a(\zeta)\Phi_i(z,\bar z)\over\zeta-
z},\label{product}\end{equation}
where the nominator is understood as OPE (\ref{nonabope}). Formula
(\ref{product}) is a definition of normal ordering for the product of
two
operators.

Let us come back to eq. (\ref{maineq}). Variation of
(\ref{maineq})
under the residual symmetry with the parameter $\epsilon_R$ 
gives rise to the following relation
\begin{equation}
\left[1-\eta\left(1-{c_V(H)\over k+2c_V(H)}\right)\right]
\partial\epsilon_R(z)
g(z)=0.\label{anomaly}\end{equation}
{}From this relation we find the renormalization constant $\eta$
\begin{equation}
\eta={k+2c_V(H)\over k+c_V(H)}.\label{eta}\end{equation}
In the classical limit $k\to\infty$, $\eta\to1$.

With the given constant $\eta$ the equation (\ref{maineq}) reads off
\begin{equation}
\left\{{\partial\over\partial z}~+~{k+2c_V(H)\over
k+c_V(H)}A(z)~+~{2\over\kappa}J(z)\right\}g(z)=0,\label{maineq'}
\end{equation}
where $A(z)$ acts on $g$ from the right hand side.
In the same fashion, the constant $\kappa$ can be calculated from
variation of
(\ref{maineq})
under the residual symmetry with the parameter $\epsilon_L$, which
leads to 
\begin{equation}
\kappa={1\over k+c_V(G)}.\label{kappa}\end{equation}
Note that the given expression for $\kappa$ is consistent with the %%@
condition that the combination $\partial+\eta A$ acted on $g$ as a Virasoro %%@
generator $L_{-1}$:
\begin{equation}
L_{-1}g={2J^A_{-1}J^A_0\over k+c_V(G)}g,\label{vir}\end{equation}
where
\begin{equation}
J^A_n=\oint{dz\over2\pi}z^nJ^A(z),\label{aff}\end{equation}
with $J^A$ being defined by eqs. (2.10).

All in all, with the regularization given by eq. (\ref{product}) and
the
Ward identity (\ref{Ward}) the eq. (\ref{maineq'}) gives rise to the
following
differential
equation
\begin{equation}
\left\{{1\over2}
{\partial\over\partial z_i}~+~\sum^N_{j\ne i}\left({t^A_it^A_j\over
k+c_V(G)}-{\bar t^a_i\bar t^a_j\over k+c_V(H)}\right){1\over z_i-
z_j}\right\}\langle\langle\Phi_1(z_1,\bar z_1)\Phi_2(z_2,\bar
z_2)\cdot\cdot\cdot\Phi_N(z_N,\bar z_N)
\rangle\rangle=0,\label{gko}\end{equation}
where $t^A_i\in{\cal G}$ and $\bar t^a_i\in{\cal H}$. 
An important check of the consistency of this equation is that it
should have a flat connection, ie. if we write (\ref{gko})
as $\partial_iG=W_iG$, the connection $W_i$ should satisfy
\beq\partial_jW_i - \partial_iW_j=[W_i, W_j].\label{flat}\eeq
Since the connection in $W_i$ in (\ref{gko}) is a sum of KZ-type
terms, (\ref{flat}) follows simply from the fact 
that $[t^a,\bar{t}^b]=0$, so the equation is indeed consistent.

Equation (\ref{gko}) is our main result. By solving it, one can find
dressed
correlation
functions in the gauged WZNW model. The solutions can be expressed as
products of the correlation functions in the WZNW model for the group
$G$ at level $k$ and $H$ at level $-2c_V(H)-k$.
In particular, for the two-point
function
the equation yields
\begin{equation}
{1\over2}\partial\langle\langle\Phi_i(z,\bar
z)\Phi_j(0)\rangle\rangle=-\left[{t^A_it^A_j\over
k+c_V(G)}~-~{\bar t^a_i \bar t^a_j\over
k+c_V(H)}\right]{1\over z}\langle\langle\Phi_i(z,\bar
z)\Phi_j(0)\rangle\rangle.\label{two}\end{equation}
By the projective symmetry, the two-point function has the following
expression
\begin{equation}
\langle\langle\Phi_i(z,\bar
z)\Phi_j(0)\rangle\rangle={G_{ij}\over|z|^{4\Delta_i}},
\label{twopoint}
\end{equation}
where $\Delta_i$ is the anomalous conformal dimension of $\Phi_i$
after the
gauge dressing and $G_{ij}$ is the Zamolodchikov metric which can be
diagonalized. After substitution of expression (\ref{twopoint}) into
eq.
(\ref{two}), and  using the fact that, as a consequence of the
residual symmetry (\ref{nonab}), the dressed correlation functions
must be singlets of both the left residual group $G$ and the right
residual group $H$, we find
\begin{equation}
\Delta_i={c_i(G)\over k+c_V(G)}~-~{c_i(H)\over
k+c_V(H)},\label{dimensions}\end{equation}
where $c_i(G)=t^A_it^A_i,~c_i(H)=\bar t^a_i \bar t^a_i$.

Up to now we have only considered simple groups, but the above
analysis can easily be extended to include semi-simple groups. Of
particular interest are $G/H$ coset models where $G = H \times H$ and
the diagonal $H$ subgroup is gauged. In this case the action
(\ref{action}) becomes
\bea
S(g,\tilde{g},A)&=&S_{WZNW}(g,k_1)~+~S_{WZNW}(\tilde{g},k_2)
\nonumber \\
&+&{k_1\over2\pi}\int d^2z\mbox{Tr}\left[
Ag^{-1}\bar\partial g - \bar A\partial g g^{-1}
+Ag^{-1}\bar Ag
{}~-~
A\bar
A\right] \\
&+&{k_2\over2\pi}\int d^2z\mbox{Tr}\left[
A\tilde{g}^{-1}\bar\partial \tilde{g} - \bar A\partial 
\tilde{g} \tilde{g}^{-1}
+A\tilde{g}^{-1}\bar A\tilde{g}
{}~-~
A\bar
A\right] \nonumber
,\label{action2}\eea
where $S_{WZNW}(g,k)$ is the same as in (\ref{wznw}), $g \in H_{k_1}$
and $\tilde{g} \in H_{k_2}$.
The equations of motion in the gauge (\ref{newgauge}) are now
\beq
\bar{\partial} J_1 =\bar{\partial} J_2 = \bar{\partial} A = 0,
\label{geqmotion2} \eeq
where,
\bea
J_1&=&-{k_1\over2}\partial
gg^{-1}~-~{k_1\over2}gAg^{-1} \nonumber \\
J_2&=&-{k_2\over2}\partial
\tilde{g}\tilde{g}^{-1}~-~{k_2\over2}\tilde{g}A\tilde{g}^{-1}
\label{J2} \eea
The residual symmetries (\ref{nonab}) take the form
\begin{eqnarray}
\tilde\delta\Phi_i&=&(\epsilon^a_L
t^a_i~+~\tilde{\epsilon}^a_L
s^a_i~+~\epsilon_R^a\{\bar t^a + \bar s^a\})\Phi_i,\nonumber\\&
\label{nonab2}
\\
\tilde\delta A&=&-\partial\epsilon_R-[\epsilon_R,
A],\nonumber\end{eqnarray}
where $t^a$ are the generators for $H_{k_1}$ and $s^a$ 
are the generators for $H_{k_2}$, so that $t^a+s^a$ are the generators
of the diagonal subgroup.

As in eq. (\ref{maineq}), eq. (\ref{J2}) can be presented in 
the  form
\begin{equation}
\frac{1}{2}\partial (g\tilde{g})~+~
\frac{\eta}{2}(g\tilde{g})A~+~\frac{1}{\kappa_1} J_1g\tilde{g}~+~
\frac{1}{\kappa_2} J_2g\tilde{g}=0.
\label{maineq2}\end{equation}
The renormalization constants $\eta$, $\kappa_1$ and $\kappa_2$
can be found in the same way as for simple groups, giving a
differential equation similar to (\ref{gko}).
\bea
\left\{{1\over2}
{\partial\over\partial z_i}~+~\sum^N_{j\ne i}\left({t^a_it^a_j\over
k_1+c_V(H)}+{s^a_is^a_j\over
k_2+c_V(H)}-
{(\bar t^a_i+\bar s^a_i)(\bar t^a_j+\bar s^a_j)\over k_1+k_2+c_V(H)}
\right){1\over z_i-
z_j}\right\}\times&& \nonumber \\
\times\langle\langle\Phi_1(z_1,\bar z_1)
\cdot\cdot\cdot\Phi_N(z_N,\bar z_N)
\rangle\rangle=0.&&\label{gko2}\eea
Note that we have derived an equation (\ref{gko}) or (\ref{gko2}) for
the holomorphic part of the correlation function only. This is because
we started from the gauge (\ref{newgauge}). We could also have used
the gauge $A=0$, in which case we would have derived the same equation
with $z$ replaced by $\bar z$, and $t^a$  by $\bar t^a$. Therefore for
gauge invariant correlation functions the holomorphic and
antiholomorphic parts will be a
solutions to (\ref{gko}).

\section{Minimal Models}

To illustrate the use of (and as a check of the correctness of) 
equation (\ref{gko2}), we consider the unitary minimal series of
models, with central charge 
$c=1-\frac{6}{m(m+1)}$, which is well known to coincide with the coset %%@
model
for $\frac{SU(2)_k \times SU(2)_1}{SU(2)_{k+1}}$, with $m=k+2$. 
The dimensions of
primary fields in these models are given by the Kac formula, which can
be presented in the form 
\beq
h_{r,s} = \frac{[r(m+1)-sm]^2-1}{4m(m+1)} 
= \frac{j(j+1)}{k+2} - \frac{j'(j'+1)}{k+3} + 
\frac{\epsilon(\epsilon+1)}{3} + n^2
\label{hrs} \eeq
Here $r=2j+1$, $s=2j'+1$, $\epsilon=0$ and $n^2=(j-j')^2$ if $j-j'$ is an 
integer, and 
$\epsilon=1/2$ and $n^2=(j-j')^2-\frac{1}{4}$ if $j-j'$ is a half
integer. $r,s$ are integers with $1 \le r\le m$ and $1 \le s\le m+1$.

We shall construct fields in the minimal model from primary
fields of the WZNW model as follows. We observe
 that the primary field $\phi_j$ with isospin
 $j$ under $SU(2)_k$ and a singlet under $SU(2)_1$, will have isospin
$j$ in $SU(2)_{k+1}$ and so, according to eq.(\ref{dimensions}) will have %%@
dimension
\beq
h_j = \frac{j(j+1)}{k+2} - \frac{j(j+1)}{k+3} = h_{r,r} .
\eeq    
The primary fields $\phi_j \times \tilde{g}$, with isospin $j$ under 
$SU(2)_k$ and isospin $1/2$ under $SU(2)_1$, have two components, with
isospin $j \pm 1/2$ in $SU(2)_{k+1}$, giving the dimensions
\beq
h_{j\pm} =  \frac{j(j+1)}{k+2} - 
\frac{(j\pm\frac{1}{2})((j\pm\frac{1}{2})+1)}{k+3} +\frac{1}{4}
=h_{r,r\pm1}
\eeq
We can therefore identify the fields $\Phi_{r,r}$ and $\Phi_{r,r\pm1}$
as
\bea
\Phi_{r,r} &=& Tr\phi_j \nonumber \\
\Phi_{r,r\pm1} &=& Tr\{\phi_j \times \tilde{g}\}_{j\pm\frac{1}{2}}
\label{phi=} \eea
In the WZNW model for $SU(2)_k$ the primary fields have isospins 
$0, \frac{1}{2}, \ldots, \frac{k}{2}$, giving 
$1\le r\le m$ and $1 \le s\le m+1$ as expected. For $SU(2)_1$ there
is only $j=0$ or $\frac12$, so we cannot construct any other fields in
this way. 
 The differential equation (\ref{gko2}) for a correlation
function of these fields is 
\bea
\left\{{1\over2}
{\partial\over\partial z_i}~+~\sum^N_{j\ne i}\left({t^A_it^A_j\over
k+2}+{s^A_is^A_j\over
3}-{(\bar t^A_i+\bar s^A_i)(\bar t^A_j+\bar s^A_j)
\over k+3}\right){1\over z_i-
z_j}\right\}\times \label{minimaleq} \\
\langle\langle\Phi_1
(z_1,\bar z_1)\cdots
\Phi_N
(z_N,\bar z_N)
\rangle\rangle=0,\nonumber\eea
$t^A$ and $s^A$ are the generators of the 
$SU(2)_k$ and $SU(2)_1$ respectively. All the solutions to 
eq. (\ref{minimaleq})   can be written as  products of
correlation functions of primary fields in the WZNW model at levels
$k$, $1$, and $(-k-5)$ (the $(-k-5)$ factor comes from writing 
$-\frac{1}{k+3}=\frac{1}{(-k-5)+2}$);
\beq
\langle\langle\Phi_1(z_1,\bar z_1)\cdots
\Phi_N(z_N,\bar z_N)
\rangle\rangle= 
f_k(z_1,\dots,z_N)
f_1(z_1,\dots,z_N)f_{-k-5}(z_1,\dots,z_N)
\eeq
where $f_k(z_1,\dots,z_N)$ is a solution to the KZ equation 
for $SU(2)_k$:
\bea
\left\{{1\over2}
{\partial\over\partial z_i}~+~\frac{1}{z_i-z_j}
\sum^N_{j\ne i}{t^A_it^A_j\over k+2}\right\}
f_k(z_1,\dots,z_N)&=&0 \nonumber \\
\left\{{1\over2}
{\partial\over\partial z_i}~+~\frac{1}{z_i-z_j}
\sum^N_{j\ne i}{s^A_is^A_j\over3}\right\}
f_1(z_1,\dots,z_N)&=&0 \nonumber \\
\left\{{1\over2}
{\partial\over\partial z_i}~-~\frac{1}{z_i-z_j}
\sum^N_{j\ne
i}{(\bar{t}^A_i+\bar{s}^A_i)(\bar{t}^A_j+\bar{s}^A_j)
\over k+3}\right\}
f_{-k-5}(z_1,\dots,z_N)&=&0 \label{factorise}\eea
It was already shown in \cite{gawedzki} that the coset model
factorises into $SU(2)_k$, $SU(2)_1$ and $SU(2)_{-k-5}$ sectors.

The simplest correlation functions to calculate in this way
are the functions for the fields $\Phi_{1,2}$, $\Phi_{2,1}$ and 
$\Phi_{2,2}$. The 4-point functions for these fields can be built up
from the conformal blocks in the $SU(2)$ WZNW model for the field in
the fundamental representation of  $SU(2)$. These are \cite{Knizhnik}:
\bea
\langle g_{\epsilon_1 \bar{\epsilon}_1}(z_1, \bar{z}_1)
g^{\dagger}_{\bar{\epsilon_2} \epsilon_2}(z_2, \bar{z}_2)g_{\epsilon_3
\bar{\epsilon}_3}(z_3,\bar{z}_3)g^{\dagger}_{\bar{\epsilon_4} %%@
\epsilon_4}(z_4,
\bar{z}_4)  \rangle 
&=& \frac{1}{|z_{14}z_{23}|^{4\Delta}} 
\sum_{A,B = 0,1}G_{AB} (x,\bar{x})I_A \bar{I}_B \nonumber \\
x=\frac{(z_1-z_2)(z_3-z_4)}{(z_1-z_4)(z_3-z_2)}
~~~~~~~\Delta = \frac{3}{4(2 +k)} &&
I_1 = \delta_{\epsilon_1 \epsilon_2} \delta_{\epsilon_3 \epsilon_4}  ~~~~~
I_2 = \delta_{\epsilon_1 \epsilon_4} \delta_{\epsilon_2 \epsilon_3}  
\nonumber \eea
\bea
G_{AB} (x, \bar{x}) &=& \sum_{p,q = 0,1} U_{pq} F_{A,[k]}^{(p)} (x) 
F_{B,[k]}^{(q)}
(\bar{x}) \nonumber \\
F_{1,[k]}^{(0)} (x) &=& x^{-\frac{3}{2(2+k)}} (1-x)^{\frac{1}{2(2+k)}}
F(\frac{1}{2+k}, -\frac{1}{2+k}; \frac{k}{2+k}; x) \nonumber \\
F_{2,[k]}^{(0)} (x) &=& \frac{1}{k} x^{\frac{1+2k}{2(2+k)}} 
(1-x)^{\frac{1}{2(2+k)}}
F(\frac{1+k}{2+k}, \frac{3+k}{2+k}; \frac{2+2k}{2+k}; x) \nonumber \\
F_{1,[k]}^{(1)} (x) &=& x^{\frac{1}{2(2+k)}} (1-x)^{\frac{1}{2(2+k)}}
F(\frac{1}{2+k}, \frac{3}{2+k}; \frac{4+k}{2+k}; x) \nonumber \\
F_{2,[k]}^{(1)} (x) &=& -2 x^{\frac{1}{2(2+k)}} (1-x)^{\frac{1}{2(2+k)}}
F(\frac{1}{2+k}, \frac{3}{2+k}; \frac{2}{2+k}; x) \nonumber \\
U_{10} &=& U_{01} = 0 ~~~ U_{11} = h U_{00} \nonumber \\
h &=& \frac{1}{4} \frac{\Gamma (\frac{1}{2+k}) \Gamma
(\frac{3}{2+k})}{\Gamma (\frac{1+k}{2+k}) \Gamma (\frac{-1+k}{2+k})}
\frac{\Gamma^2 (\frac{k}{2+k})}{\Gamma^2 (\frac{2}{2+k})}
\label{blocks} \eea
In the case of $k=1$, The conformal blocks $F^{(1)}_{A,[1]}(x)$ are
excluded, and the remaining functions reduce to
\bea
F_{1,[1]} (x) &=& \left( \frac{1-x}{x} \right)^{\frac{1}{2}} 
\nonumber \\
F_{2,[1]} (x) &=& \left( \frac{x}{1-x} \right)^{\frac{1}{2}}
\label{k=1}\eea
The first example we consider is the four-point function of 
the $\Phi_{1,2}$ field. By equation (\ref{phi=}), 
$\Phi_{1,2}=\tilde{g}_{\epsilon\bar{\epsilon}}
\delta_{\epsilon\bar{\epsilon}}$,
and so we find from equations (\ref{minimaleq}) and 
(\ref{blocks}) the following conformal blocks:
\bea
(z_{14}z_{23})^{-2h_{1,2}}{\cal F}_{(1,2)}^{(p)}(x) &\sim&
\langle\langle\Phi_{1,2}(z_1)\Phi_{1,2}(z_2)
\Phi_{1,2}(z_3)\Phi_{1,2}(z_4)\rangle\rangle 
 \nonumber \\
{\cal F}_{(1,2)}^{(p)}(x) &=&
2F^{(p)}_{1,[-k-5]}(x)F_{1,[1]}(x) +
2F^{(p)}_{2,[-k-5]}F_{2,[1]}(x) + \nonumber \\
&&+F^{(p)}_{1,[-k-5]}(x)F_{2,[1]}(x)  +
F^{(p)}_{2,[-k-5]}(x)F_{1,[1]}(x) 
\label{1,2blocks}\eea
The expression for the conformal blocks for the fields 
$\Phi_{2,1}=g_{\epsilon_1\bar{\epsilon}_1}
\tilde{g}_{\epsilon_2\bar{\epsilon}_2}
\delta_{\epsilon_1\epsilon_2}
\delta_{\bar{\epsilon}_1\bar{\epsilon}_2}$ and 
$\Phi_{2,2}=g_{\epsilon\bar{\epsilon}}
\delta_{\epsilon\bar{\epsilon}}$,
are similar:
\bea
(z_{14}z_{23})^{-2h_{2,1}}{\cal F}_{(2,1)}^{(p)}(x) &\sim&
\langle\langle\Phi_{2,1}(z_1)\Phi_{2,1}(z_2)
\Phi_{2,1}(z_3)\Phi_{2,1}(z_4)\rangle\rangle 
 \nonumber \\
{\cal F}_{(2,1)}^{(p)}(x) &=&
2F^{(p)}_{1,[k]}(x)F_{1,[1]}(x) +
2F^{(p)}_{2,[k]}F_{2,[1]}(x) + \nonumber \\
&&+F^{(p)}_{1,[k]}(x)F_{2,[1]}(x)  +
F^{(p)}_{2,[k]}(x)F_{1,[1]}(x) \nonumber \\
(z_{14}z_{23})^{-2h_{2,2}}{\cal F}_{(2,2)}^{(p,q)}(x) &\sim&
\langle\langle\Phi_{2,2}(z_1)\Phi_{2,2}(z_2)
\Phi_{2,2}(z_3)\Phi_{2,2}(z_4)\rangle\rangle 
 \nonumber \\
{\cal F}_{(2,2)}^{(p,q)}(x) &=&
2F^{(p)}_{1,[k]}(x)F^{(q)}_{1,[-k-5]}(x) +
2F^{(p)}_{2,[k]}F^{(q)}_{2,[-k-5]}(x) + \nonumber \\
&&+F^{(p)}_{1,[k]}(x)F^{(q)}_{2,[-k-5]}(x)  +
F^{(p)}_{2,[k]}(x)F^{(q)}_{1,[-k-5]}(x) 
\label{2,blocks}\eea
This gives  two solutions for the  conformal blocks for the
$\Phi_{1,2}$ and $\Phi_{2,1}$ fields, and four for the  $\Phi_{2,2}$
field, but the equation (\ref{minimaleq}) is actually a fourth
order
differential equation in all three cases, and so there are two more
solutions
(which are given by replacing $F_{A,[1]}(x)$ by $F_{A,[1]}^{(1)}(x)$
 in (\ref{1,2blocks}) and (\ref{2,blocks})). When the $\bar{x}$ 
dependence is restored by
imposing conditions of crossing symmetry and locality, the two extra
solutions do not contribute to the full 4-point function.
Using eq. (\ref{k=1}), the conformal blocks for the 
$\Phi_{1,2}$ and $\Phi_{2,1}$ fields
can be reduced to expressions containing only one hypergeometric
function, which agree with the conformal blocks given in
\cite{dotsenko}. 
\bea
{\cal F}_{(1,2)}^{(0)}(x) &=&
2x^{-2h_{1,2}}(1-x)^{h_{1,3}-2h_{1,2}}
F\left(\frac{k+2}{k+3},\frac{1}{k+3};\frac{2}{k+3};x\right)
\nonumber \\
{\cal F}_{(1,2)}^{(1)}(x) &=& 
\frac{3k}{k+1}[x(1-x)]^{h_{1,3}-2h_{1,2}}
F\left(\frac{k+2}{k+3},\frac{2k+3}{k+3};\frac{2k+4}{k+3};x\right)
\nonumber \\
{\cal F}_{(2,1)}^{(0)}(x) &=& 
 2x^{-2h_{2,1}}(1-x)^{h_{3,1}-2h_{2,1}}
F\left(\frac{k+3}{k+2},\frac{-1}{k+2};\frac{-2}{k+2};x\right)
\nonumber \\
{\cal F}_{(2,1)}^{(1)}(x) &=& 
 \frac{3(k+5)}{k+4}[x(1-x)]^{h_{3,1}-2h_{2,1}}
F\left(\frac{k+3}{k+2},\frac{2k+7}{k+2};\frac{2k+6}{k+2};x\right)
\label{minblocks}\eea

The trace in eq. (\ref{phi=}) ensures that all fields are gauge
singlets. If instead we simply considered  $g_{\epsilon,\bar
{\epsilon}}$ or $\tilde{g}_{\epsilon,\bar
{\epsilon}}$, we would find similar solutions to (\ref{minimaleq}) but
there are no simultaneous non-trivial solutions 
to the equation for the $\bar x$
dependence.
In this way we can express all correlation functions of 
$\Phi_{r,r}$ fields in terms of correlation functions in $SU(2)_k$ 
and   $SU(2)_{-k-5}$ WZNW models, and functions of $\Phi_{r,r\pm1}$
 in terms of functions in $SU(2)_k$, $SU(2)_1$ 
and   $SU(2)_{-k-5}$ WZNW models. Equation (\ref{minimaleq}) 
does not apply to fields with $|r-s|>1$, since in deriving
eq. (\ref{gko2}) we assumed that we were dealing with primary fields
of the ungauged WZNW model. However, using the relation
$h_{r,s}=h_{m-r,m+1-s}$, fields with $s=r+2$ can also be
considered as fields with $r-s=1$,
ie. $\Phi_{r,r+2}=\Phi_{m-r,m-r-1}$, and so we can also use
(\ref{minimaleq}) to find correlation functions involving these
fields.  It can be seen from eq. (\ref{hrs}) that the fields which
obey eq. (\ref{minimaleq}) include all the relevant fields
($h_{r,s}<1$) in the unitary minimal models. 
Although we cannot compute all correlation functions directly from
(\ref{minimaleq}), that equation nevertheless does in principle contain
complete information about all correlation functions, as all the
fields $\phi_{r,s}$ can be obtained from the operator product
expansion of fields with $|r-s| \le 1$. For example, using the OPE 
\beq
\Phi_{2,1} \times \Phi_{2,1} \sim [I] + [\Phi_{3,1}],
\label{ope} \eeq
which can be deduced from (\ref{minblocks}), we can obtain the n-point
function of $\Phi_{3,1}$ fields from the 2n-point function of
$\Phi_{2,1}$ fields. 

\section{$SU(2)/U(1)$ and $SL(2,{\cal R})/U(1)$ Models}

In this section we discuss the use of eq.(\ref{gko}) in the closely
related coset models for $SU(2)/U(1)$ and $SL(2,{\cal R})/U(1)$. In
both cases the central charge is
\beq c=\frac{3k}{k+2}-1, \eeq
but in the case of 
$SU(2)/U(1)$ $k$ is a positive integer, while in $SL(2,{\cal R})/U(1)$
$k$ is negative and not necessarily an integer. In particular,
$k=-9/4$, $c=26$ gives the 2D back hole \cite{Witten1}. The  
$SU(2)/U(1)$ model with integer $k$ is the ${\cal Z}_k$ parafermion
model of \cite{Zamolodchikov}.
Our convention for the
sign of $k$ in  $SL(2,{\cal R})/U(1)$ is opposite to  \cite{Witten1},
so that we can use the same equations for both $SU(2)/U(1)$ 
and $SL(2,{\cal R})/U(1)$.
Also, in  $SL(2,{\cal R})/U(1)$ the
$U(1)$ subgroup can be either compact or non-compact. In the case
when $U(1)$
is compact, our equation takes the following form
\begin{equation}
\left\{
{\partial\over\partial z_i}~+~2\sum^N_{j\ne i}\left({t^A_it^A_j\over
k+2}-{\bar t^3_i\bar t^3_j\over k}\right){1\over z_i-
z_j}\right\}\langle\langle\Phi_1(z_1,\bar z_1)\Phi_2(z_2,\bar
z_2)\cdot\cdot\cdot\Phi_N(z_N,\bar z_N)
\rangle\rangle=0,\label{compact}
\end{equation}
where $t^A_i\in {\cal SL}(2)$. While in the non-compact case, the
equation is
\begin{equation}
\left\{
{\partial\over\partial z_i}~+~2\sum^N_{j\ne i}\left({t^A_it^A_j\over
k+2}+{\bar t^3_i\bar t^3_j\over k}\right){1\over z_i-
z_j}\right\}\langle\langle\Phi_1(z_1,\bar z_1)\Phi_2(z_2,\bar
z_2)\cdot\cdot\cdot\Phi_N(z_N,\bar z_N)
\rangle\rangle=0,\label{noncompact}
\end{equation}
in the case of $SU(2)/U(1)$, the $U(1)$ subgroup is always compact,
and so the equation is (\ref{compact}) with $t^A_i\in {\cal SU}(2)$.
The conformal dimension given by  eq. (\ref{dimensions}) for a field
$\Phi_{j,m}^{j,\bar m}$, with $SU(2)$ or $SL(2,{\cal R})$ isospin $j$
and $t^3 \Phi_{j,m}^{j,\bar m}=m\Phi_{j,m}^{j,\bar m}$, $\bar{t}^3
\Phi_{j,m}^{j,\bar m}=\bar{m}\Phi_{j,m}^{j,\bar m}$ is
\begin{equation}
\Delta^{\bar m}_j = \frac{j(j+1)}{k+2} - g_{33} \frac{\bar m^2}{k}.
\label{sl2dims} \end{equation}
In the case of a compact $U(1)$, $g_{33} = +1$, and in the
non-compact case $g_{33} = -1$. In the case of $SU(2)$, $j$ is of
course an integer or half-integer with $0 \le j \le k/2$, and,
$m=-j,-j+1,\dots,j$.  However, for the  $SL(2,{\cal R})/U(1)$ 
model of $2D$ black holes \cite{Witten1} we are most
interested in the infinite dimensional representations of 
$SL(2,{\cal R})$, and so $j$ can take any value. If the $U(1)$ group
is non-compact, 
correlation functions will only be gauge invariant, (and satisfy the
differential equations for  both $z$ and $\bar z$) if $m+\bar m = 0$,
while if the $U(1)$ group is compact, we can have  $m+\bar m = nk$ for
integer $n$. Of course, in the case of $SU(2)/U(1)$ $|m|\le k/2$
anyway, so we can only have $n=0$.

As with the minimal models, the solutions to 
 eqs. (\ref{compact}) and (\ref{noncompact}) can be expressed 
as products of conformal blocks  from 
$SU(2)_k$ or $SL(2,{\cal R})_k$ and $U(1)_{-k}$ WZNW models, giving:
\beq 
\langle\langle\Phi_{j_1,m_1}^{j_1,\bar m_1}(z_1)\dots
\Phi_{j_N,m_N}^{j_N,\bar m_N}(z_N)\rangle\rangle \sim
\prod_{i<j}(z_i-z_j)^{\frac{2g_{33}\bar{m}_i\bar{m}_j}{k}}
G_{m_1,\dots,m_N}(z_1,\dots,z_N).
\label{sl2xu1}\eeq
Where $G_{m_1,\dots,m_N}(z_1,\dots,z_N)$ is a solution to the KZ
equation for $SU(2)_k$ or $SL(2,{\cal R})_k$, and the prefactor is  a 
$U(1)_{-k}$ correlation function.
 The simplest example is the
four-point function for the field in the fundamental
 representation of  $SU(2)$, $\Phi = 
\Phi_{\frac12,\frac12}^{\frac12,-\frac12} + 
\Phi_{\frac12,-\frac12}^{\frac12,\frac12}$.
In order to use (\ref{sl2xu1}) it is convenient to write 
the  $SU(2)_k$ conformal blocks
(\ref{blocks}) as 
\bea
F^{(a)}_{+-+-}(x) &=& F^{(a)}_{1,[k]}(x) \nonumber \\
F^{(a)}_{++--}(x) &=& F^{(a)}_{2,[k]}(x) \nonumber \\
F^{(a)}_{+--+}(x) &=& -F^{(a)}_{1,[k]}(x)-F^{(a)}_{2,[k]}(x). 
\label{blocks'}\eea
Using eqs. (\ref{blocks}), (\ref{sl2xu1}) and  (\ref{blocks'}), the
conformal blocks in the $SU(2)/U(1)$ model can now be written as :
\bea
{\cal F}^{(0,1)}(x) &=&
x^{-2\Delta} (1-x)^{\Delta_1-2\Delta}
F(\frac{1}{2+k}, -\frac{1}{2+k}; \frac{k}{2+k}; x) \nonumber \\
{\cal F}^{(0,2)}(x) &=&
\frac{1}{k} x^{\Delta_2-2\Delta} 
(1-x)^{\Delta_3-2\Delta}
F(\frac{1+k}{2+k}, \frac{3+k}{2+k}; \frac{2+2k}{2+k}; x) \nonumber \\
{\cal F}^{(1,1)}(x) &=&
x^{\Delta_3-2\Delta} (1-x)^{\Delta_1-2\Delta}
F(\frac{1}{2+k}, \frac{3}{2+k}; \frac{4+k}{2+k}; x) \nonumber \\
{\cal F}^{(1,2)}(x) &=&
-2 x^{\Delta_1-2\Delta} (1-x)^{\Delta_3-2\Delta}
F(\frac{1}{2+k}, \frac{3}{2+k}; \frac{2}{2+k}; x) \nonumber \\
\label{parablocks}\eea
where $\Delta_1=\frac{2}{k+2}-\frac{1}{k}=\Delta_1^1$,
$\Delta_2=1-\frac{1}{k}=\Delta_{k/2}^{k/2-1}$,
and $\Delta_3=\frac{2}{k+2}=\Delta_1^0$. When $k=1$, $c=0$ and
$\Delta=0$, and (\ref{parablocks}) simplifies to 
${\cal F}^{(0,1)}(x)={\cal F}^{(0,2)}(x)=1$. When $k=2$, $c=\frac12$,
 (\ref{parablocks}) reduces to the conformal blocks for the
$\Phi_{1,2}$ field for the minimal model with $k=1$ (\ref{minblocks}).
The full four-point function
must obey both eq. (\ref{compact}) and, because of gauge invariance, 
 the same equation with $z_i$,
$t_i^A$ and $\bar{t}_i^3$ replaced by $\bar{z}_i$, $\bar{t}_i^A$ and
$t^3_i$ respectively. The general solution is:
\beq
\langle\langle \Phi(z_1)\Phi(z_2)\Phi(z_3)\Phi(z_4)
\rangle\rangle = (z_{14}z_{23})^{-2\Delta}\sum_{a,b=0}^1\left[
U^1_{a,b}{\cal F}^{(a,1)}(x){\cal F}^{(b,1)}(\bar{x}) +
U^2_{a,b}{\cal F}^{(a,2)}(x){\cal F}^{(b,2)}(\bar{x})\right].
\label{gensol} \eeq
The constraints of locality at $x=0,1$ and crossing symmetry imply
\beq
U^A_{0,1}=^A_{1,0}=0,~~~~~~
U^A_{1,1}=hU^A_{0,0},~~~~~~
U^1_{0,0}=U^2_{0,0}, ~~~~~~A=1,2
\eeq
$h$ is the same as in eq. (\ref{blocks}). Eq. (\ref{gensol}) does not
contain terms such as 
${\cal F}^{(a,1)}(x){\cal F}^{(a,2)}(\bar{x})$, even
before locality and crossing symmetry are imposed,  
because this would be a solution to eq. (\ref{compact})
for the function 
$\langle\langle\Phi_+^-(1)\Phi_-^+(2)\Phi_+^-(3)\Phi_-^+(4)
\rangle\rangle$, and to
the equation for $\bar{x}$ for the different function 
$\langle\langle\Phi_+^-(1)\Phi_+^-(2)\Phi_-^+(3)\Phi_-^+(4)
\rangle\rangle$.

The solution above also applies in the case of 
$SL(2,{\cal R})/U(1)$ with a compact $U(1)$, but correlation functions
of fields in finite dimensional representations of $SL(2,{\cal R})$ are
not very
useful for the study of the 2D black hole.
In order to study the infinite dimensional representations of the coset %%@
$SL(2)/U(1)$, we make use of the following representation of the $SL(2)$ %%@
generators:
\begin{eqnarray}
t^+&=&{\partial\over\partial y},\nonumber\\
t^-&=&y^2{\partial\over\partial y}~-~2jy,\label{generators}\\
t^3&=&y{\partial\over\partial y}~-~j.\nonumber\end{eqnarray}
Then, the residual symmetries impose the following constraints
\begin{equation}
\sum^N_{n=1}\left[y^{l+1}_n{\partial\over\partial y_n}~-%%@
~(l+1)jy^l_n\right]\langle\langle\Phi_1(z_1,y_1)...\Phi_2(z_2,y_2)
\rangle\rangle=0,\end{equation}
where $l=-1,~0,~1$.
These constraints are nothing but the Ward identities of the $SL(2)$ %%@
projective conformal group acting on parameters $y_n$. In particular, the %%@
four-point function depending on $y_1,~y_2,~y_3,~y_4$, due to the Ward %%@
identities, can be presented as follows
\begin{equation}
G(y_1,~y_2,~y_3,~y_4)=\left[(y_1-y_4)(y_2-%%@
y_3)\right]^{2j}F(t),\end{equation}
where
\begin{equation}
t={(y_1-y_2)(y_3-y_4)\over(y_1-y_4)(y_3-y_2)},\end{equation}
the function $F$ depends only on $t$ , whereas $j$ is the $SL(2)$ spin of %%@
the operator $\Phi$.
With the representation (\ref{generators}), eq. (\ref{compact}) takes the
form
\bea
&&\left\{\frac12\frac{\partial}{\partial z_i} +
\sum_{j\neq i}^N \frac{1}{z_i-z_j}\left[\frac{1}{2(k+2)}
\left(-(y_i-y_j)^2
\frac{\partial^2}{\partial y_i\partial y_j} -
2(y_i-y_j)\left(j_j\frac{\partial}{\partial y_i}-
j_i\frac{\partial}{\partial y_j}\right) + 2j_ij_j \right) 
\right.\right. \nonumber \\
&&\left.\left.
-\frac{1}{k}\left(\bar{y}_i\bar{y}_j
\frac{\partial^2}{\partial \bar{y}_i\partial \bar{y}_j}
-j_i\bar{y}_j\frac{\partial}{\partial \bar{y}_j}
-j_j\bar{y}_i\frac{\partial}{\partial \bar{y}_i}
+j_ij_j\right)
\right]\right\}
\langle\langle\Phi_1(1)\cdots\Phi_N(N)\rangle\rangle=0. 
\label{sl2eqn}\eea
Here $j_i$ is the $SL(2)$ spin of the operator $\Phi_i$,
$\Phi(1)=\Phi(z_1,\bar{z}_1,y_1,\bar{y}_1)$ and we have
used the representation (\ref{generators}) for both left and right
$SL(2)$ groups, so $\bar{t}^3$ becomes
$\bar{y}\frac{\partial}{\partial \bar{y}}-j$. As long as we are only
interested in the $z$ and not the $\bar{z}$ dependence of the
correlation functions, we can replace $\bar{t}^3_i$ by the $U(1)$ charge
$\bar{m}_i$. The solutions to eq. (\ref{sl2eqn}) can then, as before, be
written as products of $SL(2)$ and $U(1)$ factors:
\beq
\langle\langle\Phi_1^{\bar{m}_1}(1)\cdots\Phi_N^{\bar{m}_N}(N)
\rangle\rangle\sim F_{SL(2)}(z_1,y_1,\dots,z_N,y_N)
\prod_{i<j}(z_i-z_j)^{\frac{2g_{33}\bar{m}_i\bar{m}_j}{k}}
\eeq
Where  $F_{SL(2)}(z_1,y_1,\dots,z_N,y_N)$ satisfies
\beq
\left\{(k+2)\frac{\partial}{\partial z_i} +
\sum_{j\neq i}^N \frac{1}{z_i-z_j}
\left(-(y_i-y_j)^2
\frac{\partial^2}{\partial y_i\partial y_j} -
2(y_i-y_j)\left(j_j\frac{\partial}{\partial y_i}-
j_i\frac{\partial}{\partial y_j}\right) + 2j_ij_j 
\right)\right\}
F_{SL(2)}=0
\label{sl2grav}\eeq
Equation (\ref{sl2grav}) is  closely related to the
equation for correlation functions in gravitationally dressed CFTs
derived in \cite{bilal}. Indeed, if we put $j_i=-\Delta$, 
$k+2=-\gamma$, $z_i=x_i^+$  and
$y_i=z_i^-$ in (\ref{sl2grav}), we get equation (2.41) from
\cite{bilal}\footnote{In eqs (\ref{sl2eqn}) and (\ref{sl2grav}) we
have used $t^At^A \equiv \eta_{\alpha \beta}t^{\alpha}t^{\beta}$, with
$\eta_{33} = 1$, $\eta_{+-}=\eta_{-+}=-1/2$. Since $k$ is negative,
this gives the $SL(2, {\cal R})$ metric $k\eta_{\alpha \beta}$ the
signature $(++-)$.}. The important difference is that in the case of 
gravitationally dressed CFTs both the coordinates appearing in 
(\ref{sl2grav}) are world-sheet coordinates, while in the present case
$y$ is an extra $SL(2)$ coordinate. Another unusual feature of
eq.(\ref{sl2eqn}) is that the fields $\Phi_i$ each contain a number of
different primary fields (because  $\Phi_i$ has components with
different $U(1)$ charges.
To see this, we write $\Phi_i$ as
\beq
\Phi(z,\bar{z},y,\bar{y})=\sum_{m,\bar{m}}
\Phi_m^{\bar{m}}(z,\bar{z})y^{j+m}\bar{y}^{j+\bar{m}}.
\label{phiexp}\eeq
The solution to (\ref{sl2eqn}) for the 2-point function is then:
\beq
\langle\langle\Phi_m^{\bar{m}}(z,\bar{z})\Phi_{-m}^{-\bar{m}}(0) 
\rangle\rangle \propto \frac{1}{z^{2\Delta_j^{\bar{m}}}}
\eeq
We now turn to the four-point function of the fields $\Phi$. As above,
the conformal blocks can be written as products of $SL(2)_k$ and
$U(1)_{-k}$ factors, and eq. (\ref{sl2grav}) for the 
the $SL(2)$ factor becomes
\bea
F_{SL(2)} = [(y_1-y_3)(y_2-y_4)]^{-2j}[(z_1-z_3)(z_2-z_4)
]^{-\frac{2j(j+1)}{k+2}}G(x,t) \nonumber \\
\left\{(k+2)x\frac{\partial}{\partial x} - 
\frac{1-t}{1-x}(x-t)\frac{\partial}{\partial t}t
\frac{\partial}{\partial t} - 
(1+4j)t\frac{\partial}{\partial t} - 
2j^2\frac{t+1}{t-1}\right\}G(x,t)=0
\label{partialeq}\eea
Unlike the cases we have considered up to now, this 
 partial differential equation is difficult to solve exactly. In
\cite{bilal}, a solution was found for  the behaviour as $t\goto 1$, 
but to find the behaviour of four point functions of the primary fields
$\Phi_m^{-m}$ we need to know the behaviour 
 to all orders in $t$, even if we are
only interested in the region where $x \goto 0$ or $x \goto 1$. Even
so, the solution from \cite{bilal} is interesting because it has a
logarithmic dependence on $t$ and $x$  which presumably will also
occur in the full solution. The solution to leading order in $t-1$ is:
\beq
G(x,t)\sim \left[\frac{t-1}{\log x}\right]^{-2j}
\left[2\psi(1)-\psi(-2j)-\log (-k-2) -
\log \left(\frac{t-1}{\log x}\right) +\cdots \right]
\eeq
The $\log (t-1)$ term indicates the appearance in the operator product
expansion of an operator ${\cal O}$ for which, instead of
(\ref{phiexp}), we have the expansion
\beq
{\cal O}(z,y)=\sum_m\left({\cal O}_m(z)+
\tilde{\cal O}_m(z)\log y\right)y^{j+m}
\label{logO}\eeq
It can be seen from (\ref{generators}) that ${\cal O}_m(z)$
and $\tilde{\cal O}_m(z)$ form a representation in which $t^3$ cannot
be diagonalized, and has the Jordan block structure:
\bea  
t^3\tilde{\cal O}_m(z) &=& m\tilde{\cal O}_m(z) \nonumber \\
t^3{\cal O}_m(z) &=& m{\cal O}_m(z) + \tilde{\cal O}_m(z)
\label{Jblocks}\eea
This is similar to the situation when a four point function has a
logarithmic dependence on $x$, which indicates the appearance in the
OPE of logarithmic operators, which form Jordan blocks for the Virasoro
generator $L_0$ \cite{gur}. This is not surprising, as it was
predicted in \cite{kogmav} that logarithmic operators should 
 appear in the spectrum of the $2D$ black hole. However, the operators
${\cal O}_m(z)$ which exist in the $SL(2,{\cal R})/U(1)$ coset model
and in gravitationally dressed CFT, and also presumably in other 
 coset models based on non-compact groups, differ from the
logarithmic operators that have been studied up to now in that they form
indecomposable representations of both the Kac-Moody and Virasoro
algebras. In an ordinary logarithmic CFT, there is a pair of operators
$C$ and $D$ which satisfy:
\bea
L_0C &=& \Delta C \nonumber \\
L_0D &=& \Delta D + C
\label{CD}\eea 
In the $SL(2,{\cal R})$ WZNW model, the operator ${\cal O}(z,y)$,
although it is in a reducible but indecomposable representation of
$SL(2,{\cal R})$, is an ordinary primary field of the Virasoro
algebra, since using eqs. (\ref{logO}) and (\ref{generators}) together
with $J^A_0 {\cal O}(z,y) = t^A {\cal O}(z,y)$ we find as usual:
\beq
L_0^{(WZNW)}{\cal O}(z,y) =\frac{1}{k+2}J^A_0J^A_0{\cal O}(z,y) =
\frac{j(j+1)}{k+2}{\cal O}(z,y)
\eeq
However, in the coset model we have:
\beq
L_0{\cal O} = \frac{1}{k+2}t^A_0t^A_0{\cal O} -
 \frac{1}{k}\bar{t}^3_0\bar{t}^3_0{\cal O}
\eeq
which follows from (\ref{two}) (the analysis leading to
eq. (\ref{gko}) is the same for the operators ${\cal O}$ as for
ordinary operators). Using eq. (\ref{Jblocks}) we now find:
\bea
L_0 \tilde{\cal O}_m &=& \left(\frac{j(j+1)}{k+2} -
\frac{\bar{m}^2}{k}\right) \tilde{\cal O}_m \nonumber \\
L_0 {\cal O}_m &=& \left(\frac{j(j+1)}{k+2} - \frac{\bar{m}^2}{k}\right)
{\cal O}_m - \frac{2\bar{m}}{k}\tilde{\cal O}_m
\eea
If we now put $C=-\frac{2\bar{m}}{k}\tilde{\cal O}_m$ and $D={\cal O}_m$,
this takes the usual form (\ref{CD}). We can therefore see that
operators which form indecomposable representations of $SL(2,{\cal R})$
are primary operators in the WZNW model but become logarithmic
operators after the model is gauged.

\section{Mix of gravitational and gauge dressings}

One can consider a gauged WZNW model coupled to 2D gravity. As is well %%@
known 2D gravity affects both anomalous conformal dimensions  and %%@
correlation functions \cite{kpz}. Therefore, it is interesting to study the %%@
gravitational effects on the gauged WZNW models. Our starting point, as %%@
usual, will be the classical equation of motion of the gauged WZNW model.

In the presence of 2D gravity taken in the light cone gauge \cite{polyakov}
\begin{equation}
\bar h=0,\end{equation}
the equation of motion is given as follows
\begin{equation}
-{k\over2}\left[\partial~+~h\bar\partial~+~\bar\Delta(\bar\partial %%@
h)\right]g~-~{k\over2}gA={\cal J}g,\end{equation}
where $\bar\Delta$ is the conformal dimension of $g$, classically, %%@
$\bar\Delta=0$. Here ${\cal J}$ is an affine current which obeys the same %%@
commutation relations as $J$ in eqs. (2.10).

At the quantum level due to renormalization of the singular products, the %%@
equation of motion becomes
\begin{equation}
\left[\partial~+~h\bar\partial~+~\bar\Delta(\bar\partial h)\right]g~-~\eta %%@
gA~+~{2\over\kappa}{\cal J}g=0.\end{equation}
Because in the light cone gauge 2D gravity does not change the residual %%@
symmetries, the constants $\eta,~\kappa$ will be the same as without %%@
gravity,
\begin{equation}
\eta={k+2c_V(H)\over k+c_V(H)},~~~~~~~~\kappa={1\over %%@
k+c_V(G)}.\end{equation}
Whereas the dressed conformal dimension $\bar\Delta$  is to be found from %%@
the gravitational Ward identities combined with the gauge Ward identities.

The gravitational Ward identities universal for all CFT's have been derived %%@
in \cite{polyakov}. In particular,
\begin{eqnarray}
&&\ll h(z)\Phi_1(z_1,\bar z_1)....\Phi_N(z_N,\bar z_N)\gg\nonumber\\ & & \\
&=&-{1\over\gamma}
\sum^N_{i=1}\left[{(\bar z-\bar z_i)^2\over z-z_i}{\partial\over\partial %%@
\bar z_i}~-~2\bar\Delta_i{\bar z-\bar z_i\over z-z_i}\right]\ll %%@
\Phi_1(z_1,\bar z_1)....\Phi_N(z_1,\bar z_N)\gg.\nonumber \end{eqnarray}
Here the constant $\gamma$ is given as follows
\begin{equation}
\gamma\equiv K+2={c-13\pm\sqrt{(c-1)(c-25)}\over12},\end{equation}
where
\begin{equation}
c={k\dim G\over k+c_V(G)}~-~{k\dim H\over k+c_V(H)},\end{equation}
and the +sign is chosen for $c\geq25$, and the $-$sign for $c\leq1$.

By using this identity and its derivative with respect to $\bar z$ in %%@
combination with the gauge Ward identities, one can finally arrive at the %%@
following equation
\begin{eqnarray}
&&\left\{{1\over2}{\partial\over\partial z_i}~+~\sum^N_{j\ne %%@
i}\left[{t^A_it^A_j\over k+c_V(G)}~-~{\bar t^a_i\bar t^a_j\over %%@
k+c_V(H)}\right]{1\over z_i-z_j}
~+~{1\over2\gamma}\sum^N_{j\ne i}\left[
{(\bar z_i-\bar z_j)^2\over z_i-z_j}{\partial^2\over\partial\bar %%@
z_i\partial\bar z_j}\right.\right.\nonumber\\ & & \\
&&\left.\left.~+~2\bar\Delta{\bar z_i-\bar z_j\over z_i-%%@
z_j}\left({\partial\over\partial\bar z_j}~-~{\partial\over\partial\bar %%@
z_i}\right)~-~{2\bar\Delta^2\over z_i-z_j}\right]\right\}\ll \Phi(z_1,\bar %%@
z_1)....\Phi(z_N,\bar z_N)\gg=0.\nonumber\end{eqnarray}
The latter equation can be presented as follows
\begin{equation}
\left\{{1\over2}{\partial\over\partial z_i}~+~\sum^N_{j\ne %%@
i}{{t^A_it^A_j\over k+c_V(G)}~-~{\bar t^a_i\bar t^a_j\over k+c_V(H)}
~-~{\eta_{\alpha\beta}\bar T^\alpha_i\bar T^\beta_j\over K+2}\over z_i-%%@
z_j}\right\}\ll \Phi_1(z_1,\bar z_1)....\Phi_N(z_N,\bar %%@
z_N)\gg=0,\end{equation}
where the $SL(2)$ generators $\bar T_i^\alpha$ are given as follows
\begin{equation}
\bar T_i^\alpha=\bar z_i^{\alpha+1}{\partial\over\partial\bar z_i}~-%%@
~(\alpha+1)\,\bar\Delta_i\,\bar z_i^\alpha,~~~~~~~~~\alpha=-%%@
1,~0,~+1.\end{equation}

In the case of the two-point function, one can obtain the following %%@
relation for the dressed conformal dimensions:
\begin{equation}
\Delta_i~-~\Delta_0{}_i={\bar\Delta_i(\bar\Delta_i-%%@
1)\over\gamma},\end{equation}
where $\Delta_0{}_i$ is the conformal dimension without gravity.
>From our differential equation itself, it does not follow that the %%@
holomorphic $\Delta_i$ and the antiholomorphic $\bar\Delta_i$ conformal %%@
dimensions coincide. However, Lorentz symmetry imposes one extra condition %%@
on the conformal dimensions, which is as follows
\begin{equation}
\Delta_i~-~\Delta_0{}_i=\bar\Delta_i~-~\bar\Delta_0{}_i.\end{equation}
This constraint guarantees that the Lorentz spin of operators is not %%@
affected by the gravitational dressing. Thus, our differential equation %%@
gives rise to the known KPZ formula
\begin{equation}
\bar\Delta_i~-~\bar\Delta_0{}_i={\bar\Delta_i(\bar\Delta_i-%%@
1)\over\gamma},\end{equation}

\section{Conclusion}

By using the Polyakov chiral gauge approach, we have derived a few KZ-type %%@
equations which allow us to compute correlation functions of gauged WZNW %%@
models. We have checked that our equations give the correct expressions for %%@
correlators of the minimal models. In the case of Witten's 2D black hole, %%@
we saw that our equation closely resembles the equation describing the %%@
gravitational dressing. It might be interesting to pursue further this %%@
analogy to consider the case of the continuous unitary representations of %%@
$SL(2)$. As is known these representations admit complex values for the %%@
spin $j$ \cite{bargman}. Since, following the above analogy with 2D %%@
gravity, $j$ is to be identified with the dressed conformal dimension, it %%@
is curious to know what implications would this fact mean for 2D gravity?

Another interesting question is about a relation between our equation %%@
(\ref{gko}) and the equation obtained in \cite{Halpern1} on the bases of %%@
the affine-Virasoro construction. At least for the minimal models both the %%@
equations give rise to the correct correlation functions. At the same time, %%@
it is obvious that the two equations have different structures. Indeed, the %%@
equation in \cite{Halpern1} has its connection $W_i$ already expressed in %%@
terms of some correlation functions. We can think of one possible %%@
explanation to the existence of two different equations having one and the %%@
same solutions. Namely, this situation may occur in gauge invariant %%@
theories when one uses two different gauges. However, we do not know what %%@
gauge could lead to the equation in \cite{Halpern1}, because it is not %%@
clear to us how this equation can be derived directly from the equations of %%@
motion of the gauged WZNW model. We also would like to point out other %%@
generalizations of the KZ equation considered in \cite{alekseev}

The equations which we have studied in the present paper have many more %%@
solutions than we have actually discussed. Some of them may be important %%@
for a better understanding of the dynamical properties of the gauged WZNW %%@
models. Especially, it is interesting to look for solutions to eq. %%@
(\ref{sl2grav}) corresponding to the continuous representations as they %%@
appear to be very significant for 2D black holes.

\section{Acknowledgments}

We have benefited from discussions with  J. Cardy, N. Mavromatos  and
 especially A. Tsvelik.

\end{document}